\documentclass[%
    secnumarabic,
    superscriptaddress,
    preprint,
    preprintnumbers,
    amsmath,amssymb,
    showkeys
    ]{revtex4-2}

\usepackage{graphicx}
\usepackage{dcolumn}
\usepackage{bm}
\usepackage{hyperref}
\usepackage[mathlines]{lineno}

\usepackage{todonotes}
\usepackage{verbatim}
\usepackage{placeins}

\usepackage{layouts}
\begin{document}


\title{Evidence-based policy-making in sports funding using a data-driven optimization approach}

\author{Jan Hurt}
\affiliation{Complexity Science Hub Vienna, Vienna, Austria}

\author{Liuhuaying Yang}
\affiliation{Complexity Science Hub Vienna, Vienna, Austria}

\author{Johannes Sorger}
\affiliation{Complexity Science Hub Vienna, Vienna, Austria}

\author{Thomas J.~Lampoltshammer}
\affiliation{University for Continuing Education Krems, Krems, Austria}

\author{Nike Pulda}
\noaffiliation

\author{Ursula Rosenbichler}
\noaffiliation

\author{Stefan Thurner}
\affiliation{Section for Science of Complex Systems, CeMSIIS, Medical University of Vienna, Vienna, Austria}%
\affiliation{Complexity Science Hub Vienna, Vienna, Austria}
\affiliation{Santa Fe Institute, Santa Fe, NM, USA}

\author{Peter Klimek}
\affiliation{Section for Science of Complex Systems, CeMSIIS, Medical University of Vienna, Vienna, Austria}%
\affiliation{Complexity Science Hub Vienna, Vienna, Austria}

\date{\today}

\begin{abstract}
Many European countries face rising obesity rates among children, compounded by decreased opportunities for sports activities during the COVID-19 pandemic. Access to sports facilities depends on multiple factors, such as geographic location, proximity to population centers, budgetary constraints, and other socio-economic covariates. Here we show how an optimal allocation of government funds towards sports facilitators (e.g. sports clubs) can be achieved in a data-driven simulation model that maximizes children's access to sports facilities. We compile a dataset for all 1,854 football clubs in Austria, including estimates for their budget, geolocation, tally, and the age profile of their members. We find a characteristic sub-linear relationship between the number of active club members and the budget, which depends on the socio-economic conditions of the clubs' municipality. 
In the model, where we assume this relationship to be causal, we evaluate different funding strategies. We show that an optimization strategy where funds are distributed based on regional socio-economic characteristics and club budgets outperforms a naive approach by up to 117\% in attracting children to sports clubs for 5 million Euros of additional funding.
Our results suggest that the impact of public funding strategies can be substantially increased by tailoring them to regional socio-economic characteristics in an evidence-based and individualized way. 
\end{abstract}
\keywords{sports funding, sports policy, funding distribution optimisation, evidence based policy}

\maketitle

\section{Introduction}

Physical activity of children has many physiological and psychological benefits, ranging from decreased levels of  obesity~\cite{waters2011interventions}, improved emotional development~\cite{baciu2015quality}, increased self esteem, to reductions in back injuries~\cite{sothern1999health}. Advantages are reported in several domains of health~\cite{archer2014health}, such as the reduction of physical stress~\cite{garcia2012exercise}, benefits for brain health~\cite{sibley2003relationship}, and improved cognitive functioning~\cite{sibley2003relationship}. However, over 80\% of the world's adolescent population is insufficiently physically active, exercising less than one hour per day~\cite{who2020physicalactivity,guthold2020global}. Accordingly, the WHO member states agreed to reduce relative physical inactivity by 15\% by 2030 with respect to levels of 2016~\cite{who2019global}. 
For example, Austrian schools usually only demand 2-4 hours of exercise every week~\cite{turn_stundenzahl}.
To reach the goal of 1 hour per day, a significant amount of exercise must occur in extracurricular after-school activities. Access to physical activities has been reduced during the Covid-19 pandemic leading to an increase in obesity~\cite{lockdown_obesity_china,weight_gain_lockdown,katsoulis2021obesity,mulugeta2021impact}.

Sport clubs play a key role in increasing extracurricular physical activity. For instance, in Austria 40\% of children exercise multiple times per week in sports facilities organised by clubs~\cite{statistikaustria2020clubs}. In Germany, the majority of the population between 4-17 years regularly exercises in sports clubs~\cite{jekauc2013physical}. Many countries subsidize voluntary sports clubs to increase physical activity in the child and adolescent population~\cite{public_subsidies}, in line with the WHO recommendations~\cite{who2020guidelines}.

Increasing the percentage of the physically active population not only improves the health of individuals but  also leads to reduced costs in  the health care system~\cite{inactivity_canada,carlson2015inadequate}.

Previous studies reported a positive correlation between the amount of public subsidies to sports clubs and participation rates. For example, as a result of municipal sports subsidies in the Netherlands participation rates among the 6-17-year-old individuals increased from 75\% to 80\% for households, of average income; for low-income households the increase was even more pronounced, at almost 15\%  \cite{providing_rich}.
Another study that supports this finding shows that the correlation appears to be driven by the supply of facilities \cite{elmose2020public,bergsgard2019national}.
\begin{figure}[htbp!]
    \centering
    \includegraphics*[width=.6\columnwidth]{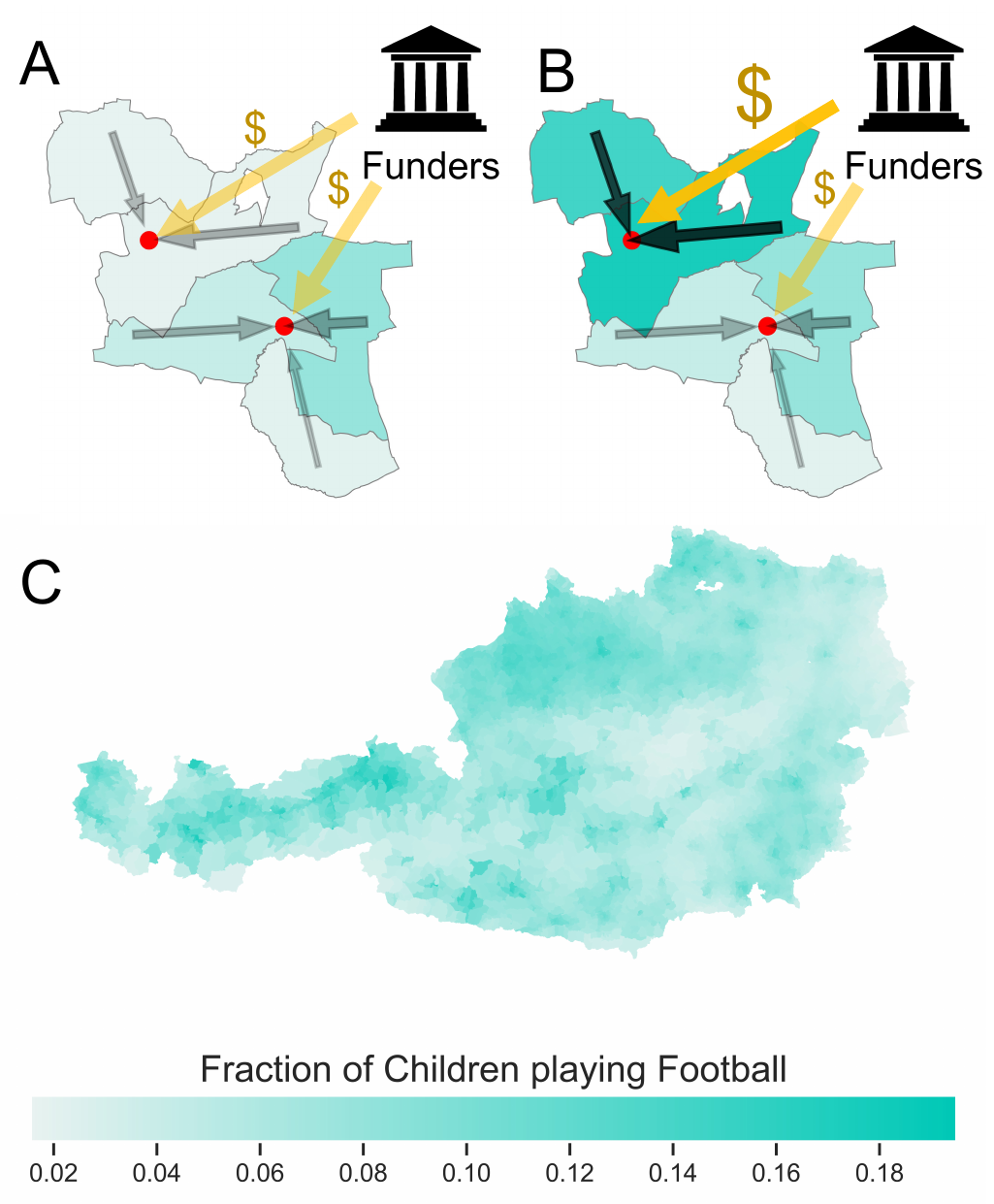}
    \caption{(A) Schematic description of the model. Funding agencies distribute funds (yellow arrows marked ``\$'') to regional sports clubs (red dots). The club attracts children from the surrounding area (black arrows) depending on the amount of funding. We work at the scale of the smallest administrative units, the  {\em counting districts} (CD). The number of children attracted depends on the number of children living in the CD and the distance between the CD and the club. (B) Increasing the funding of one specific club increases its attractiveness and thus the number of children playing football in the surrounding CDs. (C) Map of Austria showing the fraction of children playing football in every CD. There is an urban-rural differential, with more children playing football in urban areas, except for Vienna (the largest city of Austria, Vienna, is located in the North Eastern area with a low indicator value) and its surroundings.}
    \label{fig: overview}
\end{figure}

In this work, we estimate the functional relation between regional sports funding and the participation rates of children in Austrian football clubs. Assuming this relation to be causal, it becomes possible to derive an optimal funding scheme that drastically outperforms standard funding strategies, such as all distributing funds equally across clubs or proportionally to club size. We illustrate the model in \autoref{fig: overview}(A)-(B) we illustrate the model. Funders distribute subsidies across clubs, and the additional funding allows clubs, to attract additional children from the surrounding municipality.

To this end, we construct a comprehensive dataset containing practically all (1,854) football clubs with a total of approximately 111,000 children. It covers the number of members aged below 18 years, the geocoded location of the pitch, and the estimated budget depending on the competition level, i.e.,  the tier within the domestic hierarchy on which the primary adult team of the club competes. 
Employing a simple gravity model, see Data and Methods for details, we estimate the number of football-playing children for all of Austria's 8,820 counting districts (CD, being the smallest administrative unit).
A map showing the fraction of children playing football in each CD, shown in panel (C) of \autoref{fig: overview} visualizes this data. 
Tese data allow us to quantify the functional dependence of active club members and budget and some socio-economic factors that influence that dependence. Assuming this relation to be causal, we extrapolate the {\em marginal activity rate} as the number of additional children a club can attract with an additional unit of subsidies.
The idea is then to find the optimal funding strategy that allocates funds such that the overall activity is maximized. The optimal scheme is that the club with the highest marginal activity rate is funded (with small units of funding) as long as it shows the highest rate. More funds will eventually lower the marginal activity rate and once another club becomes the one with the highest marginal rate - it will be the one that receives the funding units. In this manner, funds will always be redirected to the club with the highest return. This procedure is repeated until the available funds are used up. 
This optimized funding strategy is then compared to three alternative strategies in terms of absolute activity.
The first is to share funding equally (in absolute terms) between all clubs; in the second strategy, clubs receive funding in proportion to their current budgets; the third is an optimization neglecting the socio-economic effects. We present results of how the different strategies differ in the outcome. In addition we provide an interactive dashboard \url{https://vis.csh.ac.at/sports-viz/} and an online game \url{https://vis.csh.ac.at/sports-funding-game/}. 

\section{%
    Data and Methods
    }

\subsection{Population data}
Data on the Austrian population is available on three different levels, from coarsest to most granular: on 116 districts, on 2,115 municipalities, and on 8,820 counting districts (CD).
Demographic information for each CD, such as population size, sex and age distribution, and socio-economic indicators for each district are available at the Austrian National Statistical Office~\cite{statistikaustria_zaehlsprengel, statistikaustria_gemtab}.
The latter include the percentage of people with non-Austrian citizenship, the fraction of people aged 15 and above with completed secondary or tertiary education, respectively, the fraction of those in the age range 15 to 64 that are employed, the numbers of commuters, as well as the number of corporations, number of families, available jobs per capita, and household size; the latter is available on the level of municipalities. Political boundaries for CDs are available from open government data~\cite{zaehlsprengel}. The area of every CD was calculated from polygon shapes that were then used to estimate the population density. The fraction of children (5-19 years) with respect to the total population was calculated at the district level. In order to estimate the number of children in every CD, it is assumed that this fraction is constant for every CD within a specific district.

\FloatBarrier
\subsection{Clubs}
There are 160 football competitions (leagues) in Austria are classified into ten tiers. Information on all clubs and their homepages in each competition was collected from the website of the Austrian Football Association (OEFB)~\cite{oefb}. An automated web scraping algorithm applied to these sites resulted in 1,854 links to club homepages from which it was possible to retrieve the GPS coordinates of the football court and the number of players. Clubs may have multiple teams including youth teams with a name that typically signals the age range of the players \footnote{“U-XX” where “XX” is the upper bound of the age of the players in this team. For example, a “U-12” team could include players from the ages 10-12 years old.}. Since some clubs form associations for specific age groups, individual players may be listed multiple times (once for each club that is a member of the association). These players are assigned to one club of the association with a probability proportional to club size.

The data collection and cleaning process resulted in a database containing the numbers of players grouped by age, team and club. In total, the clubs reported 225,907 active players. Since we focus on recreational, non-professional sports clubs, only teams that play in the 3rd tier or lower (higher league number) are considered for the analysis. Budget information for clubs was available stratified by competition level. It was obtained from a representative survey performed by SportsEconAustria (SpEA), the Institute for Sports Economy, among all 1,854 clubs.

\FloatBarrier
\subsection{Estimating the number of active children in every CD}
The so-called {\em gravity model} was found useful in describing the influence of geographic distance in a number of applications~\cite{anderson2011gravity}, including sport club participation rates~\cite{steinmayr2011closer}. It is assumed that the probability of a child playing football in a specific club, $P\left(club\rightarrow  cd\right)$ depends on the inverse square distance between the residential CD and the club's location.
To avoid problems at a distance of zero, we added $+5$ km to all distances, $f(r) = \{ 1/(r+5 {\rm km} )^2 \,\,  if \,\, r < 50 {\rm km}; \,\, 0 \,\,  {\rm else} \}$:
\begin{equation*}
    P\left( {\rm club} \rightarrow  {\rm CD} \right)=\frac{f\left(d\left({\rm CD} ,{\rm club} \right)\right)}{\sum_{{\rm CD}} f\left(d\left({\rm CD},{\rm club}\right)\right)} \quad.
\end{equation*}
Let $n_{{\rm children}}\left({\rm club}\right)$ denote the number of children that are members of a specific club, then the total number of children who play football in a specific CD is calculated as $n_{{\rm football}\ {\rm children}}\left( {\rm CD} \right)=\sum_{{\rm club}}{n_{{\rm children}}\left({\rm club}\right)}\cdot P\left({\rm club}\rightarrow  {\rm CD}\right)$. 

\subsection{Estimating the club-size function}
For every club, estimates of the total income, $e_{{\rm income}}\left({\rm club}\right)$, of every club, and the amount of children, $n_{{\rm children}}\left({\rm club}\right)$ per club are available. We empirically validate that the functional relationship of the number of children per club as a function of its total income can be reasonably approximated using parameters $\alpha>0$ and $\beta>0$ 
\begin{equation*}
    n_{{\rm children}}\left({\rm club}\right)=\alpha\cdot\log{\left(e_{{\rm income}}\left({\rm club}\right)\right)}+\beta \quad.
\end{equation*}
We call this the {\em club-size function}.

\subsubsection{Socio-economic dependencies}
Sport club participation rates amongst children are influenced by socio-economic factors and $\alpha$ may depend on socio-economic indicators of the municipality. To find out which indicator is most influential with respect to the club-size function, $ n_{{\rm children}}\left({\rm club}\right)$, to each club we assign socio-economic indicators of the municiapality it is located in. The list of triples of the income, number of children, and value of a third specific socio-economic indicator is written as $\left(\left(e_{{\rm income}_1},n_{{\rm children_1}},ind_1\right) \right. $ ,$ \left(e_{{\rm income}_2},n_{{\rm children}_2},ind_2\right),$ $\ldots,$ $\left. \left(e_{{\rm income}_N},n_{{\rm children}_N},ind_N\right)\right)$ and $M_{ind}$ is the median of the values in $\left(ind_1,\ldots,ind_n\right)$.
For each socio-economic property, the clubs are divided into two groups that are at least as large as or smaller than the median indicator value $M_{ind}$, respectively.
For these two groups of clubs the following regression models are evaluated separately:
\begin{align}
    \forall i: ind_i\geq M_{ind} \quad &n_{{\mathrm{children}}_i}=\alpha_\geq \log\left(e_{{\rm income}_i}\right)+\beta_\geq+\epsilon_i\\
    \forall i: ind_i<M_{ind} \quad &n_{{\rm children}_i}=\alpha_< \log\left(e_{{\rm income}_i}\right)+\beta_<+\epsilon_i
    \label{eq:group regressions}
\end{align}
Here, $\alpha_\geq$, $\alpha_<$ are the regression coefficients, $\beta_\geq$,$\beta_<$ the intercepts, and $\epsilon_i$ is a random error variable. Differences between $\alpha_\geq$ and $\alpha_<$ are expressed as $Z$-scores (measuring the number of standard errors by which the measured effect size is different from zero).
To derive a socio-economically stratified club size function, we identify the indicator for which we empirically observe the largest differences in correlations between club size and income across the different strata.

The number of children a sports club is expected to attract and add to its current members is then calculated using the following formula
\begin{align*}
    n_{{\rm add.children}}\left({\rm club}\right)\left(e_{{\rm add.funding}}\right) &= \alpha \left({\rm club}\right)\cdot \log\left(e_{{\rm curr.\ income}}\left({\rm club}\right)+ e_{{\rm add.funding}}\left({\rm club}\right)\right) \\ 
    &\phantom{\alpha \left({\rm club}\right)\cdot}\qquad  - \log\left(e_{{\rm curr.}\ {\rm income}}\right)
    \quad .
\end{align*}
Here $\alpha\left({\rm club}\right)$ are the group-specific parameters, $e_{{\rm current}\ {\rm income}}$ is the current total income of the club, and $e_{{\rm add.funding}}$ is the additional funding.

\subsection{Optimal funding strategy}
An algorithm similar in spirit to gradient descent is employed to design a strategy that is optimal for a given amount of money it maximizes the expected number of additional children becoming club members.
\begin{enumerate}
    \item Calculate the current value of $n_{{\rm add.children}}({\rm club}) \, ({\rm 1.000 \, EUR})$, for each club with its current available budget.
    \item Each of the clubs with the maximum $n_{{\rm add.children}}({\rm club}) \, ({\rm 1.000 \, EUR})$ get a small amount of money $\epsilon$ EUR.
    \item Subtract $\epsilon$ EUR from the total available budget.
    \item Go to step 1. Repeat until the entire budget is spent.
\end{enumerate}
This algorithm allows us to find the funding scheme that distributes a fixed amount of funds across the clubs with the goal of maximizing the total number of additional children.

\FloatBarrier
\section{Results}
Figure \ref{fig: overview} (C), shows a map of the percentage of children playing football in every CD, according to the employed gravity model. It reveals an urban-rural differential. More children play football in urban areas than in rural ones, except for the city of Vienna and its surroundings.

\subsection{Socio-economically stratified club-size function}

We emprically assess socio-economic characteristics might impact the likelihood of clubs having more children relates to their income. In \autoref{tab: zvals}, we report the corresponding regression coefficients that encode how club size and income differ between clubs in regions with higher ($\alpha_\geq$) or lower ($\alpha_<$) indicator values.
\begin{table}[htbp!]
    \begin{tabular}{lrrr}
    \toprule
    Socioeconomic Property &     Z \;&\;   $\alpha_\geq$ \;&\;   $\alpha_<$ \\
    \hline
    Percentage of people without Austrian citizenship &  5.79 & 55.04 & 19.60 \\
    Number of families per capita & -5.74 & 21.38 & 57.19 \\
    log(population density) &  5.63 & 54.01 & 21.73 \\
    Average number of people per household & -4.79 & 28.96 & 58.24 \\
    Percentage of commuters & -4.15 & 31.59 & 57.33 \\
    Percentage, of the ages 15 to 64, employed & -3.95 & 35.69 & 59.81 \\
    Percentage of the ages 15 and above, unemployed &  3.58 & 58.58 & 36.66 \\
    Percentage of the ages 15 and above, with teritary education &  2.52 & 50.49 & 35.24 \\
    Percentage of the ages 15 and above, with secondary education & -2.14 & 38.67 & 51.41 \\
    Number of Corporations per capita & -1.92 & 42.23 & 53.97 \\
    Percentage of people working at workplaces &  1.79 & 49.37 & 37.84 \\
    Number of work places per capita & -1.59 & 44.42 & 54.21 \\
    Percentage of ages 15 and above, with sec. or ter. education & -0.98 & 45.86 & 51.89 \\
    \hline
    \end{tabular}
    \caption{For every socio-economic indicator, results for regressions for the club size function are shown. Two groups are considered, depending on whether the specific indicator of the municipality the club belongs to is larger than ($\geq$) or less than ($<$) the median of all clubs. The table shows the coefficients $\alpha_\geq$,$\alpha_<$, and their difference ($Z$-score). Indicators are ordered by the absolute value of the $Z$-score.}
    \label{tab: zvals}
\end{table}

\begin{table}[htbp!]
    \centering
\begin{tabular}{l|cccc}\toprule
 & 
 \rotatebox{0}{\shortstack{\%   without Austrian\\ citizenship}} 
 \; & \,   
 \rotatebox{0}{\shortstack{\#families\\ per capita}} 
 \; & \,
 \rotatebox{0}{\shortstack{log(pop.\\ density)}}
 \; &\,  
 \rotatebox{0}{\shortstack{\#people\\ per household}} \\
\hline

\%   without Austrian\\ citizenship
& 1.00 & -0.84 & 0.87 &  -0.68 \\[.5em]
\# families\\ per capita
& -0.84 & 1.00 & -0.80 & 0.70 \\[.5em]
log(pop.\\ density)
& 0.87 & -0.80 & 1.00 & -0.69 \\[.5em]
\# people\\ per household
& -0.68 & 0.70 & -0.69 & 1.00 \\ 
\end{tabular}
    \caption{Pairwise Pearson correlation coefficients for the four socio-economic indicators that show the strongest effect on the number of children of a sports club.}
    \label{tab: corcoef}    
\end{table}

The largest differences ($Z$-scores) are observed for the percentage of people who do not have the Austrian citizenship, as well as the number of families per capita, the log-population density and and the average number of people per household. Some of these indicators are closely related; see \autoref{tab: corcoef} where we report pairwise Pearson correlation coefficients. From those four indicators, we choose the two that have the smallest correlation, suggesting that these indeed capture independent socio-economic determinants that influence one’s likelihood to become a member of a sports club: the percentage of people who do not have the Austrian citizenship and the average number of people per household. According to these socio-economic variables, the clubs are assigned to four groups depending on whether the values of the socio-economic properties of the club are bigger or smaller than their median. 
Table \ref{tab: clubs} shows a summary of the clubs in these four groups along with information regarding the model parameter $\alpha$ in each group and the goodness of fit value $R^2$.

The larger $\alpha$ the more additional children a club is assumed to attract for a given amount of funding. We find the smallest (largest) value of $\alpha$ for clubs in regions above(below)-median numbers of families and corporations. We find good agreement between the assumed club size function and the data, as indicated by an overall $R^2$ of $0.83$.

\begin{table}
 \resizebox{\textwidth}{!}{\begin{tabular}{rr@{\hskip 1.5em}@{\extracolsep{4pt}}rrr rrr rrr rrr}
& & &
\multicolumn{2}{c}{high \% without AT cit.} 
& \multicolumn{3}{c}{high \% without AT cit.} & \multicolumn{3}{c}{low \% without AT cit.} & \multicolumn{3}{c}{low \% without AT cit.}\\
& & & \multicolumn{2}{c}{high \#people per h.h.} & \multicolumn{3}{c}{low \#people per h.h.} & \multicolumn{3}{c}{high \#people per h.h.} & \multicolumn{3}{c}{low \#people per h.h.} \\
\cline{3-5} \cline{6-8} \cline{9-11} \cline{12-14}
tier & budget [€] & clubs & children & IQR & clubs & children & IQR & clubs &  children & IQR & clubs & children & IQR \\
\hline
 3 &           484125 &     15 &    162 & 144-198 &     23 &    171 & 105-246 &      2 &     69 &  39-99 &      3 &    152 & 112-192 \\
 4 &           110026 &     39 &     92 &  61-155 &     44 &    142 &  83-228 &     14 &     93 & 37-118 &     27 &     80 &  38-118 \\
 5 &            79065 &     65 &     89 &  50-115 &     60 &    114 &  75-174 &     35 &     53 &  39-74 &     71 &     39 &   12-76 \\
 6 &            75178 &    101 &     68 &  37-104 &     82 &     71 &  24-123 &     94 &     54 &  28-82 &     96 &     34 &   10-56 \\
 7 &            45286 &     88 &     61 &   32-81 &     90 &     39 &    0-70 &    133 &     50 &  25-70 &    106 &     42 &   14-78 \\
 8 &            48385 &     87 &     39 &   15-75 &     94 &     42 &   10-85 &     80 &     36 &   9-47 &     98 &     23 &    1-48 \\
 9 &            39952 &     44 &     54 &   31-71 &     23 &     70 &  44-104 &     60 &     60 &  41-73 &     23 &     32 &    0-70 \\
10 &            34751 &      3 &     47 &   23-68 &      1 &      8 &     8-8 &      1 &     19 &  19-19 &      0 &      0 &       0 \\
\hline
\multicolumn{2}{c}{$\alpha$, $R^2$} & \multicolumn{3}{c}{$\alpha = 44.9$, $R^2 = 0.94$} & \multicolumn{3}{c}{$\alpha = 58.1$, $R^2 = 0.79$} & \multicolumn{3}{c}{$\alpha = 15.0$, $R^2 = 0.34$} &\multicolumn{3}{c}{$\alpha = 52.7$, $R^2 = 0.92$} 
\end{tabular} }
        \caption{
        For every tier of football clubs, we give the estimated budget, the number of clubs, and the number of children (along with its inter-quartile range, IQR) in each of the four groups defined as lying in an area with a high (above the median), low (below the median) number of families and corporations per capita. For every group we also present results of the regression analysis for the club-size function. For  $\alpha$ and $R^2$; see \ref{eq:group regressions}. The overall $R^2$ is $R^2=0.83$.}
        \label{tab: clubs}
\end{table}

\begin{figure}
    \centering
    \includegraphics[width = \textwidth]{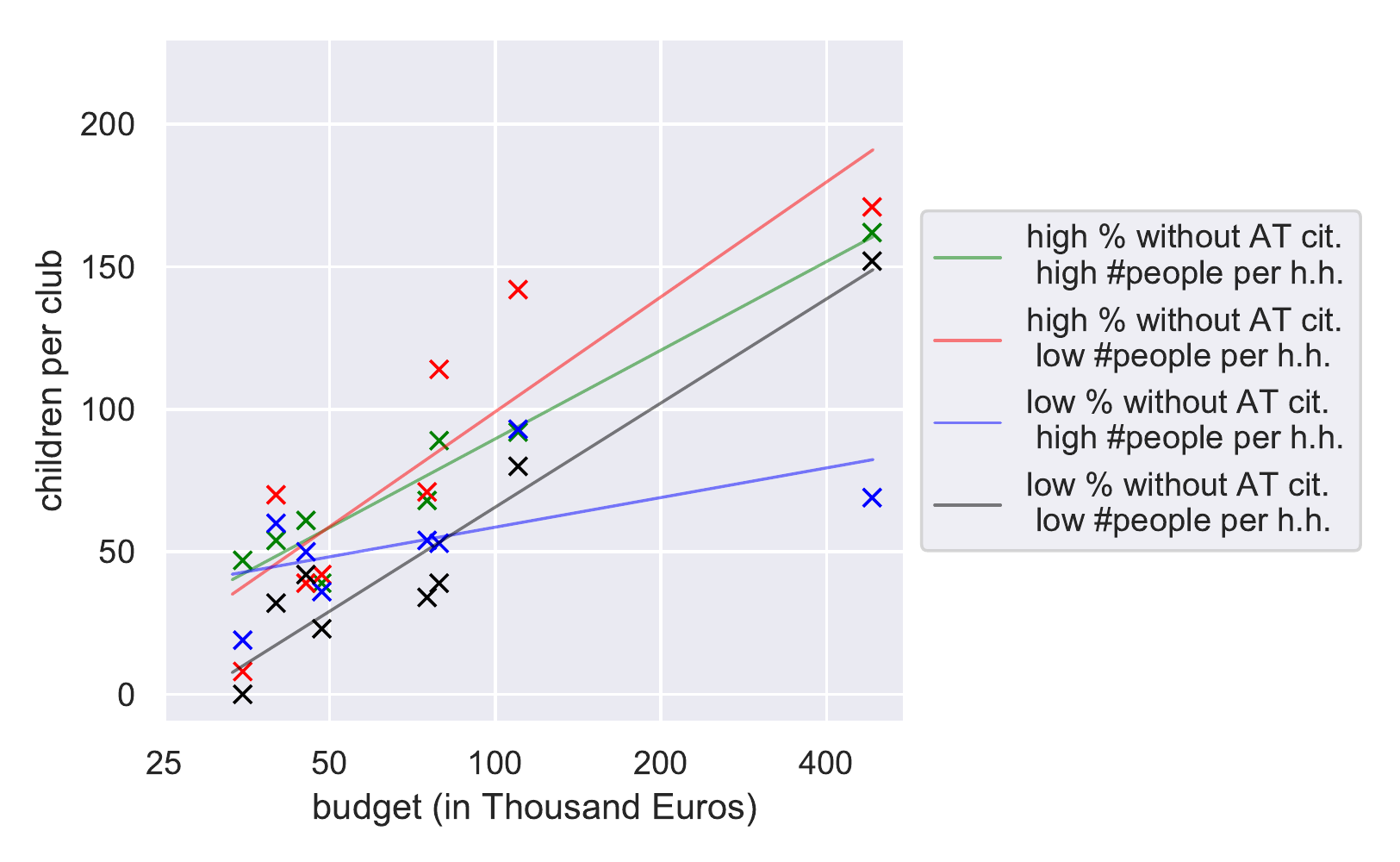}
    \caption{
    The number of children per club increases logarithmically with its budget (note the logarithmic scaling of the budget axis).
    Each cross marks the median number of children of all clubs for each competition level for each of the four groups given by high or low percentage of people who do not have the Austrian citizenship or average number of people per household in the clubs' regions.
    The slope of the fit provides the number of additional children a club in each group attracts for a unit of additional funds.
    Clearly, the slope is highest for clubs in regions with low numbers of corporations and families, meaning that a given amount of funds attracts more children than in any of the other groups.
    }
    \label{fig: club size fit}
\end{figure}

\begin{figure}
    \centering
    \includegraphics[width = 1.\columnwidth]{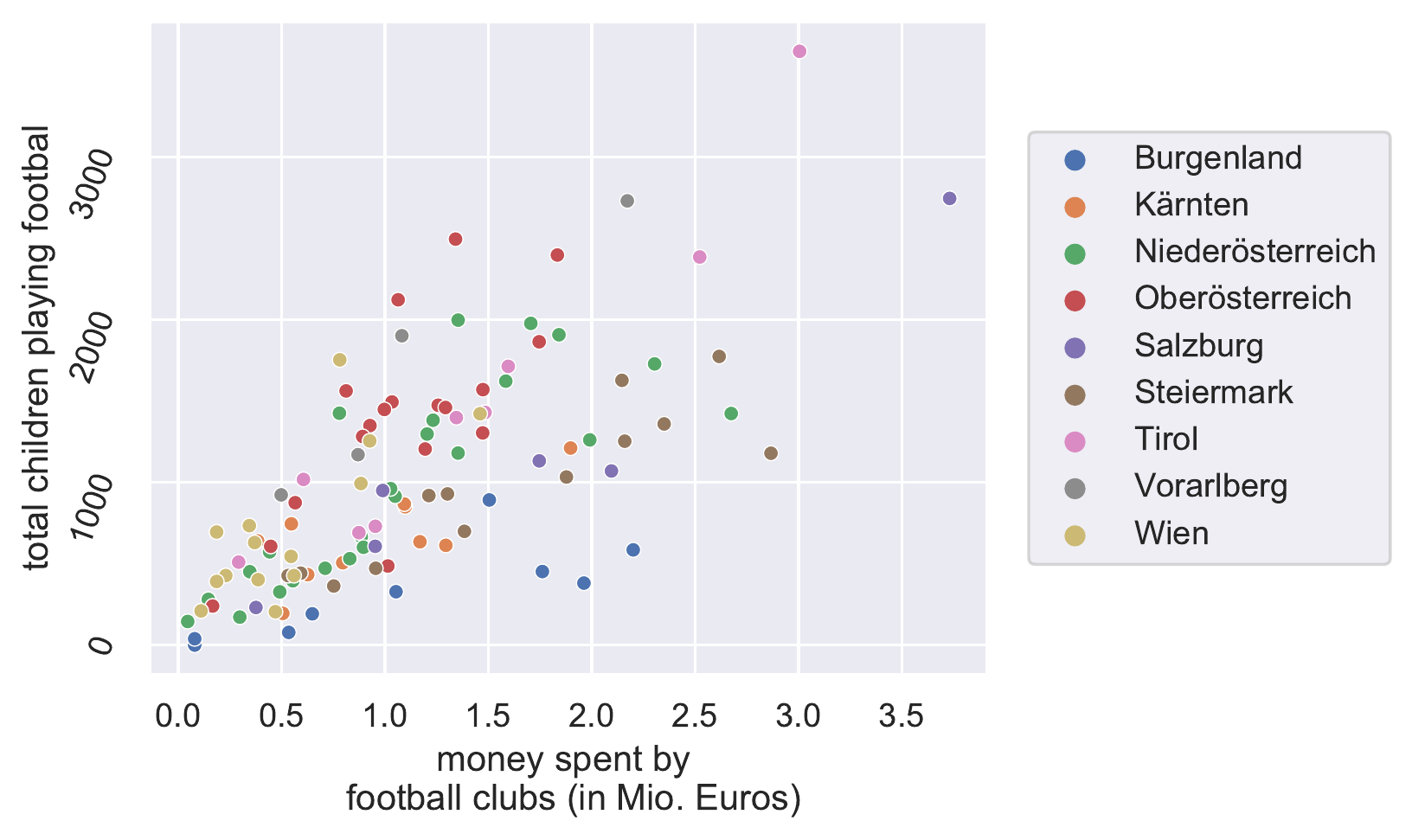}
    \caption{The number of children being members of football clubs per district increases with the total budget of all the football clubs in the district. Note the large spread in slopes for the different states of Austria. States handle funding policies autonomously.} 
    \label{fig: money children}
\end{figure}

In \autoref{fig: club size fit}, we show that the number of children per club scales logarithmically with the club's budget.
The scaling (slope) depends on regional socio-economic indicators.
We observe the steepest slope for clubs in regions with lower numbers of corporations and families compared to clubs from other regions.
\autoref{fig: money children} shows the number of children playing football as a function of money spent by the clubs for each district; colors refer to federal states. There is a large spread in slopes also across federal states.

\begin{figure}
    \centering
    \includegraphics[width = 1.\columnwidth]{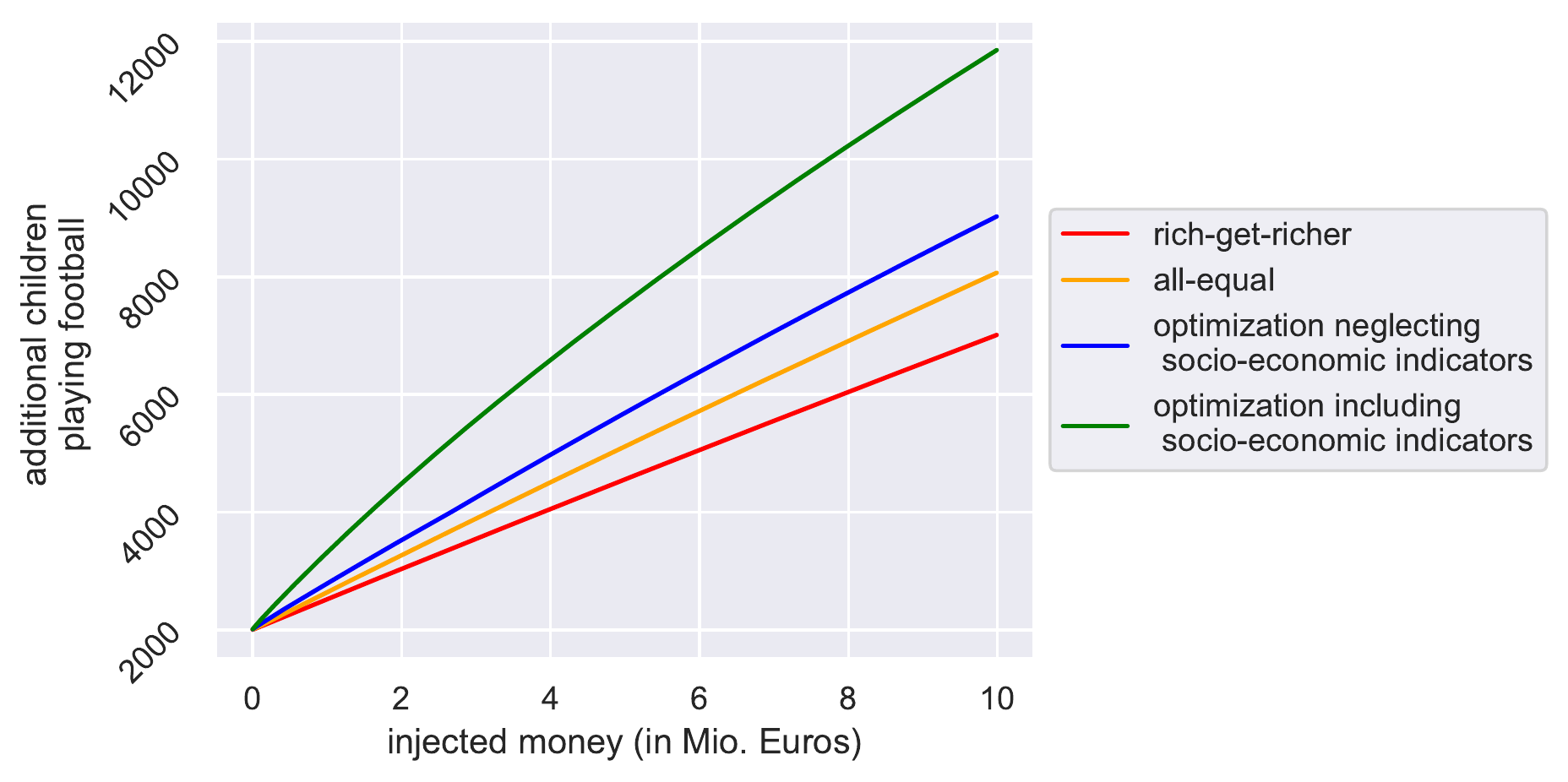}
    \caption{Comparison of the outcome (additional children playing football) of the four considered funding schemes. From best to worst performing they are {\em optimization including socioeconomic indicators} (green), {\em optimization neglecting of socio-economic indicators} (blue), {\em all-equal} and {\em rich-get-richter} (red).}
    \label{fig: optimization}
\end{figure}

The budget-activity curve allows us to benchmark different scenarios regarding the distribution of subsidies across sports clubs. As our central result, the effectiveness of the four employed funding scenarios is presented in \autoref{fig: optimization}. It shows the number of additional children playing football as a function of subsidies spent on all the clubs.
In the first scenario, referred to as {\em rich-get-richer} (red), where clubs receive additional subsidies as a percentage of their current income. In the second scenario, all clubs receive the same amount of money ({\em all-equal}, yellow).
As the so-called club-size function is logarithmic, the all-equal scenario leads to more children playing football.

These two baseline funding strategies are compared with two optimized funding allocation schemes (green and yellow). In the optimized strategy, which considers social indicators (green) the number of expected additional children playing football per invested Euro is almost twice as large as for the {\em rich-get-richer} scenario and still substantially larger than the {\em all-equal} case.
A strategy that optimizes the distribution of funds based on club budgets but neglects socio-economic differences (blue) performs better than the all-equal case but substantially worse than the optimization that includes such differences.
The scenarios shown in \autoref{fig: optimization} suggest that the larger the invested funds, the bigger the advantage of the optimization (bigger absolute difference in \autoref{fig: optimization}).
Regardless of actual numbers, the presented simulations suggest that optimized funding strategies might offer enormous advantages over traditional approaches to funding.

\begin{figure}[htbp!]
    \centering
    \includegraphics[width=.85\textwidth]{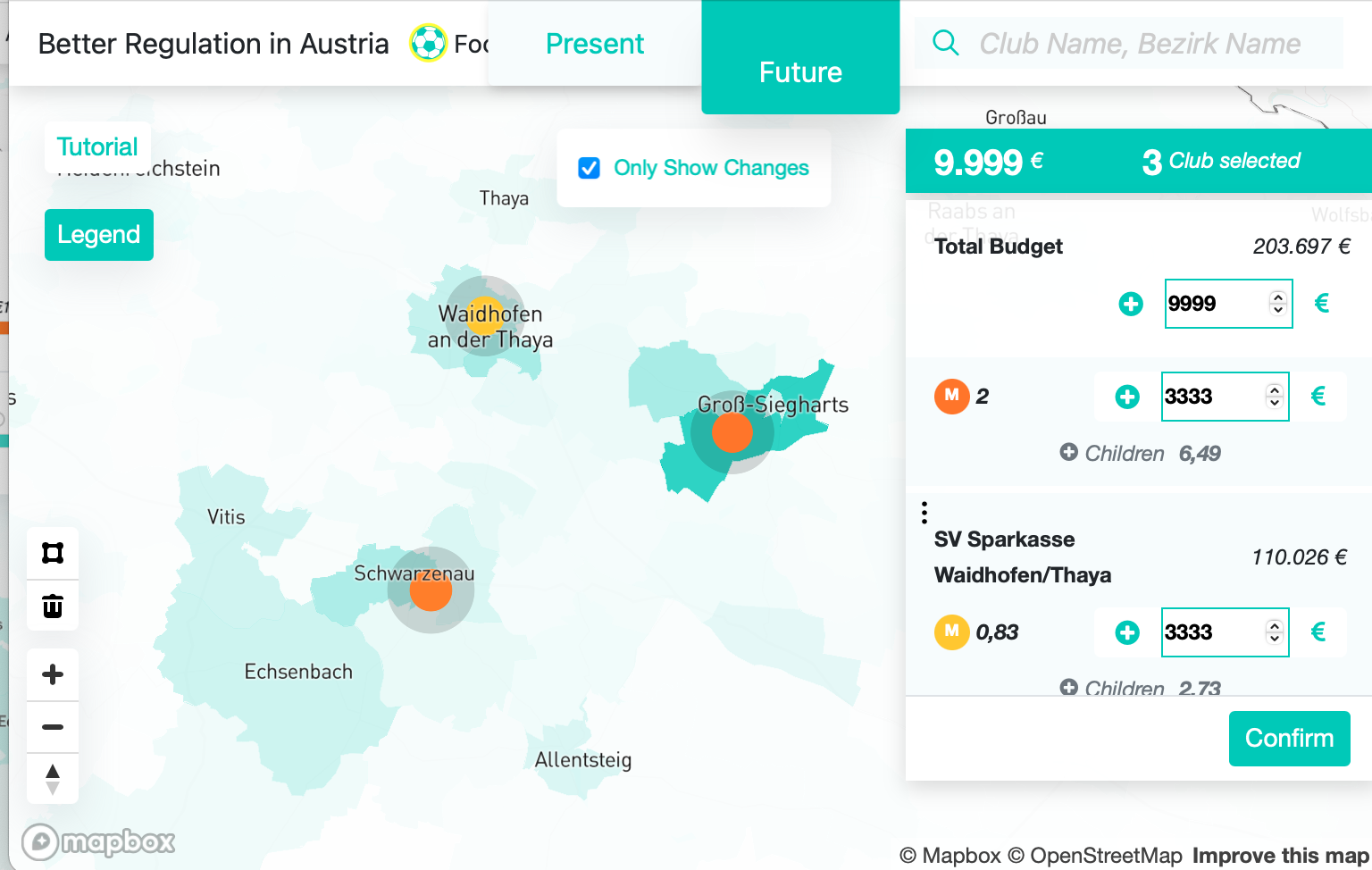}
    \caption{Dashboard for a model scenario where three clubs get subsidies of 3,333 EUR each. The football clubs are marked in yellow/orange (the color depends on the budget), and the turquoise shading of the CDs on the map indicates the expected number of additional children, and to the correct details about the clubs. On the right side panel, information on the clubs, like total budget, additional subsidies, {\em marginal activity rate} per EUR, and additional expected, children are shown.}
    \label{fig: dashboard relative changes}
\end{figure}

We provide a dashboard to allow for user-specific designs of funding strategies \footnote{\url{https://vis.csh.ac.at/sports-viz/}}, see  \autoref{fig: dashboard relative changes}.
In the dashboard, one can select sets of clubs and specify the amount of subsidies to distribute amongst those.
According to the club-size function and the gravity model, the dashboard first calculates the number of additional children for every club and then for every CD.

\FloatBarrier
\section{Discussion}
We propose a data-driven framework to identify optimal policy strategies to increase physical activity in children by increasing the number of sport club members.
We assembled a country-wide dataset of practically 1,854 non-professional football clubs, their location, competition level, and estimates for their budgets and the age information of their 225,000 members including 111,000 children. Merging this dataset with publicly available data on the Austrian population and its socio-economic status allowed us to estimate how the funding of a sports club increases the effective attractiveness for children to play at a specific club.
Compared to traditional situations in which funds are distributed equally or proportional to size, we find that  optimized strategies  substantially increase the number of children that could be attracted to join sports clubs.

The computation of an optimized funding policy requires an estimate of how funding is coupled with additional sports activity. We found a logarithmic budget--activity relation between the number of children and a club's budget, suggesting diminishing investement returns by funding larger clubs.
How fast these returns diminish is influenced by the socio-economic and structural characteristics of the club's region of the club, such as the number of families and corporations.
For a total additional funding of 5 Million Euros, the optimization neglecting socio-economic indicaotrs attracts 34\% fewer children than the optimization including socio-economic indicators.
This result demonstrates the importance of tailoring funding strategies to different socio-economic contexts.
Socio-economic factors have been found to influence expenses' effect on participation ratios~\cite{providing_rich}. Evidence shows that participation increases with household income and educational attainment~\cite{downward_2007,hovemann_2009,nobis_2019,breuer_2008}. Data from the Netherlands showed that the relationship between municipal sport expenditure and sport club participation rate showed a clear ``dose-response'' relationship with income, meaning the lower the income, the stronger the increase in participation with more funding~\cite{providing_rich}.
Together these findings suggest a ``compensation effect'' where government funding may compensate for the adverse impact of economic inequalities on physical activity.
Our results of the dependence of the budget-activity curve on socio-economic factors are in that regard in line with this literature.

In the present study we do not have access to socio-economic indicators on an individual-level. Our high spatial resolution, however, allowed us to adopt an ecological study design in which we related regional differences in socio-economic factors to the relationship between available funds (proxied by the club budget) and participation rates.
We found a particularly strong relation between club budget and participation in areas with a smaller number of corporations and work places, in line with the observed relations between economic deprivation and the impact of funding in other countries.
Our finding of a particularly strong gradient in the budget--participation relation in regions with a high number of non-Austrian citizens is of particular concern even though a strong strong correlation with the number of families per capita might confounded this relation. Further work is needed to understand how migration-related inequalities in the possibilities to access sports facilities lead to this strong gradient.

Our primary assumption in this work is that the empirical correlation captured in the budget-activity function is a causal relationship -- at least partially. A potential driver of this relationship might be that better-funded clubs can attract children without additional external funding compared to smaller clubs. The situation that larger clubs with more children attract more funding is also plausible but is not considered here.
In consequence, this means that we implicitly assume that sports participation of children is supply- and not demand-driven in the sense that if the supply of sports activities increases,  there will always be enough demand. More empirical work, ideally in the form of adequately designed surveys, would be necessary to confirm these assumptions. In the absence of a model mechanism for saturation effects, our approach should not be applied for very high participation rates close to 100\%.

In terms of data, the study is mostly limited by missing information on the exact budgets of the individual clubs.
As only competition-level estimates are available, this introduces a degree of uncertainty. Clubs often form associations to reach the number of players necessary for a team in a particular age group. As it is is impossible to reliably assign the players of these associations to particular clubs, this also introduces uncertainties in estimating regional participation rates. However, we assume that these uncertainties do not introduce systematic biases in estimating the club size functions.

Previous studies highlighted facility coverage as one of the driving factors of participation rates~\cite{elmose2020public,bergsgard2019national}.
Our data includes implicit information on the facilities via the infrastructure costs in the club's budget (football pitches, club house). It remains to be seen in future work to what extent the preseneted  results would change if club-size functions could be estimated, taking a club's available facilities into account. 

We make the presented approach more accessible to decision-makers (and the public) through an interactive dashboard that allows the user-specific design of funding policies.
Tools like this, however, are only one element among the requirements for evidence-based policy making. In addition, qualified staff is needed to analyze data~\cite{oecd_2020} and conduct plausibility checks. In the long run, in the light of evermore constrained budgets and public doubt in the efficacy of politics, it seems inevitable to integrate evidence-based policy-making into national performance management (outcome orientation). It can be valued as international best practice and displays the impact chain between long-term political goals and administrative measures on a technical level~\cite{oecd_2014,commission_2020}. 

Finally, proactive data management might require changes in existing laws that foster central databases and interoperability~\cite{commission_2017}. The communication and cooperation between politics, public administration, and science is enhanced by creating appropriate platforms for interdisciplinary exchanges.

\section*{Acknowledgements}
The project was funded by the European Union's Structural Reform Support Programme (SRSP) under Grant Agreement No. GA2020/023, and the Austrian Research Promotion Agency FFG under P 882184.

\section*{Competing Interests}
The authors declare no competing interests.

\bibliography{lit.bib}

\end{document}